\documentclass[12pt,a4paper]{article}
\makeatletter
\def\eqnarray{%
\stepcounter{equation}%
\let\@currentlabel=\theequation
\global\@eqnswtrue
\global\@eqcnt\z@
\tabskip\@centering
\let\\=\@eqncr
$$\halign to \displaywidth\bgroup\@eqnsel\hskip\@centering
$\displaystyle\tabskip\z@{##}$&\global\@eqcnt\@ne
\hfil$\displaystyle{{}##{}}$\hfil
&\global\@eqcnt\tw@$\displaystyle\tabskip\z@{##}$\hfil
\tabskip\@centering&\llap{##}\tabskip\z@\cr}
\makeatother
\newcommand{\ket}[1]{{\vert{#1}\rangle}}
\newcommand{\bra}[1]{{\langle{#1}\vert}}

\newcommand{\calh}{{\cal H}}

\newcommand{\fukuso}{{\mathbf C}}
\newcommand{\real}{{\mathbf R}}
\newcommand{\futon}{{\bf N}}

\newcommand{\wzetta}{{\vert w\vert}}
\newcommand{\zetta}{{\vert z\vert}}

\begin{document}

\title{\sl Note on Extended Coherent Operators and \\ 
Some Basic Properties}
\author{
  Kazuyuki FUJII
  \thanks{E-mail address : fujii@math.yokohama-cu.ac.jp }\\
  Department of Mathematical Sciences\\
  Yokohama City University\\
  Yokohama, 236-0027\\
  Japan
  }
\date{}
\maketitle\thispagestyle{empty}
%
%
%
%
\begin{abstract}
  This is a continuation of the paper (quant-ph/0009012). 
  In this letter we extend coherent operators and study some 
  basic properties (the disentangling formula, resolution of 
  unity, commutation relation, etc). We also propose a perspective 
  of our work.
\end{abstract}

\newpage

%
%
%
%

\section{Introduction}

This is a continuation of the paper \cite{KF1}.  For the aim of 
this letter see \cite{KF1}. 

We extend coherent operators and study some basic properties such as 
\begin{enumerate}
 \item[(a)] the disentangling formula,
 \item[(b)] the resolution of unity, 
 \item[(c)] the commutation relation,
 \item[(d)] the matrix elements,
 \item[(e)] the trace, 
 \item[(f)] the Glauber formula.
\end{enumerate}
These properties are very well known for coherent operators,  
see \cite{KF1}. We give explicit forms to (a) $\sim$ (f) for extended 
version of coherent operators (extended coherent operators 
in our terminology $\cdots$ easy and bad naming !).
 
Since our extension is in a certain sense very natural, our results must 
have been known. But the author could not find such references in spite 
of his efforts.  At any rate let us list our results.

The details and further developments of this letter will be included 
in \cite{KF2}.

\section{Coherent Operators and Basic Properties}
We make a brief review of some basic properties of coherent operators 
within our necessity. 

Let $a(a^\dagger)$ be the annihilation (creation) operator of the harmonic 
oscillator.
If we set $N\equiv a^\dagger a$ (:\ number operator), then we have 
\begin{equation}
  \label{eq:2-1}
  [N,a^\dagger]=a^\dagger\ ,\
  [N,a]=-a\ ,\
  [a^\dagger, a]=-\mathbf{1}\ .
\end{equation}
Let $\calh$ be a Fock space generated by $a$ and $a^\dagger$, and
$\{\ket{n}\vert\  n\in\futon\cup\{0\}\}$ be its basis.
The actions of $a$ and $a^\dagger$ on $\calh$ are given by
\begin{equation}
  \label{eq:2-2}
  a\ket{n} = \sqrt{n}\ket{n-1}\ ,\
  a^{\dagger}\ket{n} = \sqrt{n+1}\ket{n+1}\ ,
  N\ket{n} = n\ket{n}
\end{equation}
where $\ket{0}$ is a normalized vacuum ($a\ket{0}=0\  {\rm and}\  
\langle{0}\vert{0}\rangle = 1$). From (\ref{eq:2-2})
state $\ket{n}$ for $n \geq 1$ are given by
\begin{equation}
  \label{eq:2-3}
  \ket{n} = \frac{(a^{\dagger})^{n}}{\sqrt{n!}}\ket{0}\ .
\end{equation}
These states satisfy the orthogonality and completeness conditions
\begin{equation}
  \label{eq:2-4}
   \langle{m}\vert{n}\rangle = \delta_{mn}\ ,\quad \sum_{n=0}^{\infty}
   \ket{n}\bra{n} = \mathbf{1}\ . 
\end{equation}

\noindent{\bfseries Definition}\quad The coherent operators and coherent 
states are defined as
\begin{eqnarray}
  \label{eq:2-5-1}
      U(z) = \mbox{e}^{za^{\dagger}- \bar{z}a}
      \quad \mbox{for} \quad z \in \fukuso ,  \\
  \label{eq:2-5-2}
      \ket{z} = U(z)\ket{0}
      \quad \mbox{for} \quad z \in \fukuso .  
\end{eqnarray}

Let us list the three basic properties of coherent operators and 
coherent states, see \cite{KS} and \cite{AP} for the proofs.

\vspace{10mm}
\noindent{\bfseries (a) Disentangling Formula}\quad We have 
\begin{equation}
  \label{eq:2-6}
  U(z) = \mbox{e}^{- \zetta^{2}/2} \mbox{e}^{za^{\dagger}}
         \mbox{e}^{- \bar{z}a} 
       = \mbox{e}^{\zetta^{2}/2} \mbox{e}^{- \bar{z}a}
         \mbox{e}^{za^{\dagger}}. 
\end{equation}

\noindent
In the proof we use the elementary Baker-Campbell-Hausdorff formula
\begin{equation}
  \label{eq:2-7}
 \mbox{e}^{A+B}=\mbox{e}^{-\frac1{2}[A,B]}\mbox{e}^{A}\mbox{e}^{B}
\end{equation}
whenever $[A,[A,B]] = [B,[A,B]] = 0$, see \cite{KS}. 
This is the key formula in the theory of coherent states.

\vspace{10mm}
The coherent states are usually defined as eigenvectors of annihilation 
operator $a\ket{z} =  z\ket{z}\  \mbox{for}\  z \in \fukuso$. 
The important feature of coherent states is the following resolution  
(partition) of unity.

\noindent{\bfseries (b) Resolution of Unity}\quad We have 
 
\begin{equation}
  \label{eq:2-8}
  \int_{\fukuso} \frac{[d^{2}z]}{\pi} \ket{z}\bra{z} = 
  \sum_{n=0}^{\infty} \ket{n}\bra{n} = \mathbf{1}\ ,
\end{equation}
where we have set  $[d^{2}z] = d(\mbox{Re} z)d(\mbox{Im} z)$ for simplicity.

\vspace{10mm}

\noindent{\bfseries (c) Commutation Relation}\quad We have
\begin{equation}  
  \label{eq:2-9} 
  U(z)U(w) = \mbox{e}^{z\bar{w}-\bar{z}w}\ U(w)U(z)
\end{equation}
for  $z,\ w \in \fukuso$.

\vspace{10mm}

\noindent{\bfseries (d) Matrix Elements}\quad The matrix elements of 
$U(z)$  are  
\begin{eqnarray}
   \label{eq:2-10-1}
 &&(\mbox{i})\quad n \le m \quad 
   \bra{n}U(z)\ket{m} = \mbox{e}^{-\frac{1}{2}\zetta^2}\sqrt{\frac{n!}{m!}}
                 (-\bar{z})^{m-n}{L_n}^{(m-n)}(\zetta^2), \\
   \label{eq:2-10-2}
 &&(\mbox{ii})\quad n \geq m \quad 
   \bra{n}U(z)\ket{m} = \mbox{e}^{-\frac{1}{2}\zetta^2}\sqrt{\frac{m!}{n!}}
                 z^{n-m}{L_m}^{(n-m)}(\zetta^2),
\end{eqnarray}
where ${L_n}^{(\alpha)}$ is the associated Laguerre's polynomial defined by 
\begin{equation}
   \label{eq:2-11}
 {L_n}^{(\alpha)}(x)=\sum_{j=0}^{n}(-1)^j {{n+\alpha}\choose{n-j}}
                  \frac{x^j}{j!}. 
\end{equation}

\vspace{10mm}

\noindent{\bfseries (e) Trace}\quad We have 
\begin{equation}
   \label{eq:2-12}
    \mbox{Tr}U(z) = \pi{\delta^2}(z) \equiv \pi\delta(x)\delta(y) \quad 
    \mbox{if}\  z=x+iy.
\end{equation}

\vspace{10mm}

\noindent{\bfseries (f) Glauber Formula}\quad  Let $A$ be any observable. 
Then we have 

\begin{equation}
   \label{eq:2-13}
   A = \int_{\fukuso}\frac{[d^{2}z]}{\pi}\mbox{Tr}[AU^{\dagger}(z)]U(z)
\end{equation}
This formula plays an important role in the field of homodyne tomography.

\section{Extended Coherent Operators and Basic Properties}

We in this section define extended coherent operators and study 
the basic properties corresponding to ones of coherent operators. 

\noindent{\bfseries Definition}\quad The extended coherent operators 
and extended coherent states are defined as follows : 
\begin{eqnarray}
  \label{eq:3-1-1}
      U(z,t) &=& \mbox{e}^{za^{\dagger}- \bar{z}a + itN}
      \quad \mbox{for} \quad z \in \fukuso,\  t \in \real, \\
  \label{eq:3-1-2}
      \ket{z,t} &=& U(z,t)\ket{0}
      \quad \mbox{for} \quad z \in \fukuso,\  t \in \real .  
\end{eqnarray}

\noindent
In this definition at first sight we would want to transform 
(\ref{eq:3-1-1}) as 

\begin{equation}
  \label{eq:3-2}
   za^{\dagger}- \bar{z}a + itN = it\left(a+\frac{z}{it}\right)^{\dagger}
             \left(a+\frac{z}{it}\right)- \frac{i\zetta^2}{t}.
\end{equation}

\noindent 
But with this form it is impossible to take a limit $t \rightarrow 0$, 
so we don't use this one in this paper.
 
Let us list the corresponding properties of extended coherent operators  
(\ref{eq:3-1-1}) and extended coherent states (\ref{eq:3-1-2}).
Before stating our result let us prepare some notations.
\begin{eqnarray}
  \label{eq:3-3-1} 
     f(t) &=& \frac{\mbox{e}^{it}-1}{it}, \\
  \label{eq:3-3-2}  
     g(t) &=& \frac{\mbox{e}^{it}-(1+it)}{t^{2}}.
\end{eqnarray}
We here note that 
\begin{eqnarray}
  \label{eq:3-4-1}
     \vert f(t)\vert &=& \frac{\mbox{sin}(t/2)}{t/2}, \\
  \label{eq:3-4-2} 
     \vert f(t)\vert^{2} &=& - \left\{g(t)+ \overline{g(t)}\right\}.
\end{eqnarray}

\noindent
\begin{Large}
  Let us state our result :
\end{Large} 

\vspace{5mm}
\noindent{\bfseries (a) Disentangling Formula}\quad We have 
\begin{eqnarray}
  \label{eq:3-5-1}
  U(z,t) &=& {\mbox{e}^{g(t)\zetta^2}}
             {\mbox{e}^{f(t)za^{\dagger}}}{\mbox{e}^{itN}}
             {\mbox{e}^{-f(t)\bar{z}a}},\quad \mbox{or} \\
  \label{eq:3-5-2}
         &=& {\mbox{e}^{- \overline{g(t)} \zetta^2}}
             {\mbox{e}^{- \overline{f(t)} \bar{z}a}}{\mbox{e}^{itN}}
             {\mbox{e}^{\overline{f(t)} za^{\dagger}}}.
  \end{eqnarray}

This becomes the key formula in the theory of extended coherent states.

\vspace{10mm}
\noindent{\bfseries (b) Resolution of Unity}\quad We have 
 
\begin{equation}
  \label{eq:3-6} 
  \int_{\fukuso} \frac{\vert f(t)\vert^{2} [d^{2}z]}{\pi}\  
                 \ket{z,t}\bra{z,t} = 
  \sum_{n=0}^{\infty} \ket{n}\bra{n} = \mathbf{1}.
\end{equation}

We note that the measure to satisfy resolution of unity is not 
unique, so the following measure is not so bad :

\begin{equation}
  \label{eq:3-7}
  \int_{\real} \frac{\vert f(t)\vert^{2} \mbox{e}^{- 
                \vert t\vert}}{2} dt   
  \int_{\fukuso} \frac{[d^{2}z]}{\pi}\  \ket{z,t}\bra{z,t} = 
  \sum_{n=0}^{\infty} \ket{n}\bra{n} = \mathbf{1}\ .
\end{equation}

\vspace{10mm}

\noindent{\bfseries (c) Commutation Relation}\quad We have
\begin{equation}  
  \label{eq:3-8} 
         U(z,t)U(w,s) = \mbox{e}^{
         \left\{\overline{f(t)} \overline{f(s)}z\bar{w} 
              -  f(t)f(s)\bar{z}w \right\}
                                 }\  
         U(w\mbox{e}^{it},s)U(z\mbox{e}^{-is},t)
\end{equation}
for  $z,\ w \in \fukuso\ \mbox{and}\  t,\ s \in \real$.

\vspace{10mm}

Let us here make a change of variables $z \mapsto w = f(t)z$. Then 
matrix elements of $U(z,t)$ are written down by those of $U(w)$.

\noindent{\bfseries (d) Matrix Elements}\quad The matrix elements of 
$U(z,t)$ are 

\begin{equation}
  \label{eq:3-9} 
     \bra{n}U(z,t)\ket{m} = \mbox{e}^{
      - \left\{\frac{1}{2}+\frac{\overline{g(t)}}{\vert f(t)\vert^{2}}
                \right\}\wzetta^{2} +\ itm }\
             \bra{n}U(w)\ket{m}.
\end{equation}

\vspace{10mm}

\noindent{\bfseries (e) Trace}\quad 
For $U(z,t)$ we have an another decomposition :

\begin{equation} 
  \label{eq:3-10}
    U(z,t) = \mbox{e}^{\frac{-i}{t}\zetta^{2}}
             \mbox{e}^{\frac{i}{t}(za^{\dagger}+ \bar{z}a)}
             \mbox{e}^{itN}
             \mbox{e}^{-\frac{i}{t}(za^{\dagger}+ \bar{z}a)}, 
\end{equation}
so we have 

\begin{equation} 
  \label{eq:3-11}
   \mbox{Tr}U(z,t) = \mbox{e}^{\frac{-i}{t}\zetta^{2}}\ 
                   \mbox{Tr}\mbox{e}^{itN} =
                   \frac{\mbox{e}^{-i\zetta^{2}/t}}{1-\mbox{e}^{it}}
                   \quad (\mbox{as Abel sum}),
\end{equation}
see also (\ref{eq:3-2}).

\noindent 
We here note $\mbox{Tr}U(z,0) = \mbox{Tr}U(z) = \pi \delta^{(2)}(z)$  
, (\ref{eq:2-12}).  Then we have a natural question :

\noindent{\bfseries Question}\quad Is the following correct ?
\begin{equation} 
  \label{eq:3-12} 
       \mbox{lim}_{t \to 0}\mbox{Tr}U(z,t) = \mbox{Tr}U(z,0) 
       \quad \mbox{for} \quad z \in \fukuso.
\end{equation}

\noindent
At first sight the answer of this question seems to be no, but is yes.
This has been solved by S. Sakoda.

\vspace{10mm}

Making use of (\ref{eq:3-9}) and (\ref{eq:3-4-2})  we have the Glauber 
formula.

\noindent{\bfseries (f) Glauber Formula}\quad  Let $A$ be any observable. 
Then we have 

\begin{equation}
   \label{eq:3-13}
   A =  
       \int_{\fukuso} \frac{\vert f(t)\vert^{2}[d^{2}z]}{\pi}\ 
       \mbox{Tr}[AU^{\dagger}(z,t)]U(z,t).
\end{equation}

\vspace{10mm}
A comment is in order. When taking the limit $t \rightarrow 0$ 
these properties reduce to corresponding ones in the 
preceding section because $f(t) \rightarrow 1\  \mbox{and}\ 
g(t) \rightarrow -1/2$ from (\ref{eq:3-3-1}) and (\ref{eq:3-3-2}).

\vspace{5mm}

\section{Discussion}

In this paper we listed the six basic properties of coherent operators 
and investigated these ones for extended coherent operators.
We had a natural extension.  
The details of several caluculations performed in Section 3 will be 
published in \cite{KF2}.

Here let us give a perspective (or application) of our work. 
We can also extend the squeezed operator in \cite{KF1} as follows.

\begin{equation}
  \label{eq:4-1}
   V(z,t) = \mbox{e}^{z K_{+} - \bar{z}K_{-} + itK_{3}}
   \quad \mbox{for} \quad z \in \fukuso \ \mbox{and}\ t \in \real 
\end{equation}
where 

\begin{equation}
  \label{eq:4-2}
  K_{+}\equiv{1\over2}\left( a^{\dagger}\right)^2\ ,\
  K_{-}\equiv{1\over2}a^2\ ,\
  K_{3}\equiv{1\over2}\left( a^{\dagger}a+{1\over2}\right)\ .
\end{equation}

We have studied some basic properties of this operator in \cite{KF2},  
so that we can consider the product of two unitary operators

\begin{equation}
  \label{eq:4-3}
     U(z,t)V(w,s)\quad  \mbox{for} \quad z, w \in \fukuso\ 
                    \mbox{and}\ t, s \in \real. 
\end{equation}
It may be possible to search the same lines as in \cite{KF3} and  
\cite{MP}, \cite{GLM}.  
These will be published in an another paper, \cite{KF4}.

\vspace{5mm}
\noindent{\em Acknowledgment.}\\
The author wishes to thank S. Sakoda for his helpful comments and 
suggestions.


\end{document}